\documentclass[twocolumn,prb,showpacs]{revtex4}
\usepackage{amssymb}
\usepackage{graphicx}
\usepackage{color}

\begin{document}

\title{Spin-roton excitations in the cuprate superconductors}
\author{J. W. Mei and Z. Y. Weng}
\affiliation{Institute for Advanced Study, Tsinghua University, Beijing, 100084, China}
\date{\today }

\begin{abstract}
We identify a new kind of elementary excitations, spin-rotons, in the doped Mott
insulator. They play a central role in deciding the
superconducting transition temperature $T_{c}$, resulting in a simple $T_{c}$
formula, $k_BT_{c}\simeq E_{g}/6$, with $E_{g}$ as the characteristic
energy scale of the spin rotons. We show that the degenerate $S=1$ and $S=0$
rotons can be probed by neutron scattering and Raman scattering
measurements, respectively, in good agreement with the magnetic
resonancelike mode and the Raman \textrm{A}$_{1g}$ mode observed in the high-$%
T_{c}$ cuprates.
\end{abstract}

\maketitle

\section{introduction\label{sec:int}}

To fully understand the nature of high-$T_{c}$ superconductivity in the
cuprates, one essential task is to identify the most important
elementary excitation which controls the superconducting transition.

In a conventional BCS superconductor, the Bogoliubov quasiparticle
constitutes the most crucial low-lying excitation. In a d-wave state, nodal
quasiparticle excitations generally lead to a linear-temperature reduction
of the superfluid stiffness $\rho_s$ by\cite{LeeRMP,Lee}
\begin{equation}
\rho _{s}(T)=\rho _{s}(0)-aT  \label{eq:superfluid}
\end{equation}%
which, however, would be normally extrapolated to a transition temperature ($\rho_s(T_c)=0$)
much higher than the factual $T_{c}$ in the cuprates, based on the microwave
measurements of the penetration depth which determines the superfluid density\cite{Boyce}.

On the other hand, in view of the small superfluid density in the cuprates,
which are widely considered to be a doped Mott insulator\cite{Anderson}, the phase
fluctuation of the superconducting order parameter has been suggested\cite{Kivelson} to
play an important role in the transition regime, which can be characterized by
the following London action
\begin{equation}
L=\frac{\rho _{s}}{2}\int d^{2}\mathbf{r}(\nabla \phi +q\mathbf{A}^{e})^{2}
\label{eq:london}
\end{equation}%
where $\phi$ specifies the $U(1)$ phase of the order parameter of condensate carrying charge $q$, and $\mathbf{A}^{e}$ is the external electromagnetic field. In this point of view, the superconducting
transition is of a Kosterlitz-Thouless (KT) type\cite{Kosterlitz} with the proliferation of topological vortices%
\begin{equation}
\oint d\mathbf{r}\cdot \nabla \phi =\pm 2\pi  \label{eq:vor}
\end{equation}%
which destroy the phase coherence of superconductivity resulting in $k_{B}T_{c}\simeq \rho _{s}(T_{c}^{-})$.
\begin{figure}[h]
\includegraphics[width=\columnwidth]{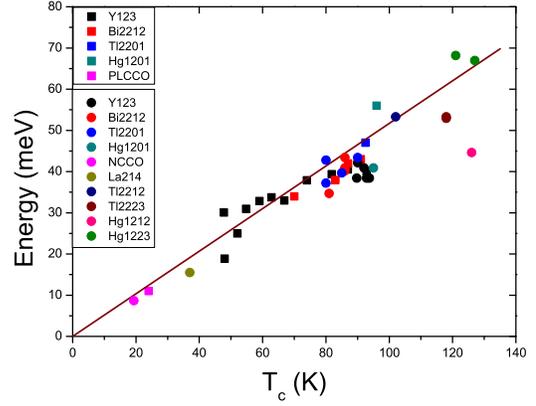}
\caption{[Color online] The characteristic energies observed by inelastic neutron
scattering (INS) and electron Raman scattering (ERS) experiments versus the superconducting
transition temperature $T_{c}$ for the high-$T_{c}$ cuprates. The
straight line shows the empirical formula (\protect \ref{eq:tcformula}), which
will be derived in the present work. Here the solid squares represent the
INS resonance mode, with different colors indicating different families
including hole-doped Y123\protect \cite{Y123}, Bi2212\protect \cite{Bi2212},
Tl2201\protect \cite{Tl2201} and Hg1201\protect \cite{Hg1201}, and electron doped $%
\text{Pr}_{0.88}\text{LaCe}_{0.12}\text{CuO}_{4-\protect \delta }$%
\protect \cite{PCCO}; the solid circles represent the ERS $A_{1g}$ mode,
including the hole doped $\text{Y}123$\protect \cite{ERS:Y123,ERS:Hg1201Y123}, $\text{Bi}2212$\protect \cite{ERS:Bi2212}, $\text{Tl}2201$%
\protect \cite{ERS:Tl2201,ERS:Tl2201Tl2212}, $\text{Hg}1201$%
\protect \cite{ERS:Hg1201Y123}, $\text{La}214$\protect \cite{ERS:La214}, $%
\text{Tl}2212$\protect \cite{ERS:Tl2201Tl2212}, $\text{Tl}2223$\protect \cite%
{ERS:Tl2223}, $\text{Hg}1212$\protect \cite{ERS:Hg1212}, $\text{Hg}1223$
\protect \cite{ERS:Hg1223} and the electron doped $\text{NCCO compound}$%
\protect \cite{ERS:NCCO}. }
\label{fig:E-Tc}
\end{figure}

However, a striking and puzzling empirical $T_{c}$ formula for the cuprate
superconductors has been known experimentally\cite{Bourges,Dai_Tc,Fong_Tc,Gallais_Tc,Uemura}, which is simply given by
\begin{equation}
k_BT_{c}=\frac{E_{g}}{\kappa }  \label{eq:tcformula}
\end{equation}%
where $\kappa\sim6$ and $E_{g}$ denotes the charcteristic
energy scales observed in inelastic neutron scattering (INS) \cite{Fong,Bourges,Dai_Tc,Fong_Tc,Y123,Bi2212,Tl2201,Hg1201,PCCO} and electronic Raman scattering (ERS) \cite{Gallais_Tc,Devereaux,ERS:Y123,ERS:Hg1201Y123,ERS:Bi2212,ERS:Tl2201,ERS:Tl2201Tl2212,ERS:La214,
ERS:Tl2223,ERS:Hg1212,ERS:Hg1223,ERS:NCCO}
measurements, as illustrated in Fig. \ref{fig:E-Tc}. Here $E_{g}$ in INS
corresponds to the well-known \textit{resonance} energy\cite{Fong} in the literature,
which is a \emph{spin-triplet} excitation at momentum centered around the AF
wave vector $\mathbf{Q}_{\text{AF}}=(\pi ,\pi )$. By contrast, $E_{g}$ in ERS corresponds to a
\emph{singlet} mode in the $\mathrm{A}_{1g}$ channel near momentum $\mathbf{Q%
}_{\text{0}}=(0,0)$. The ERS data in $B_{1g}$ and $B_{2g}$ channels
have provided the compelling evidence for the $d$-wave pairing
symmetry in the cuprate superconductors, however, the $A_{1g}$ peak at $E_{g}$ remains an unresolved mystery\cite{Devereaux}. As shown in
Fig. \ref{fig:E-Tc}, more materials can be accessible by ERS than INS,
including the $\text{La}_{2-x}\text{Sr}_x\text{Cu}\text{O}_4$ compound in which there is no direct INS
evidence for a sharp resonancelike mode  but a
singlet mode in ERS\cite{ERS:La214} has been still found with $E_{g}$ well fit by (\ref%
{eq:tcformula}).

The above empirical scaling law of $T_{c}$ vs.\ $E_{g}$ implies that the
elementary excitations controlling the superconducting transition in the
cuprates should be composed of two \emph{degenerate }modes, with quantum
number $S=0$ and $1$, respectively, as probed in ERS and INS. Note that in
the literature the magnetic resonancelike mode observed in INS has been
sometimes interpreted as the bound state of a Bogoliubov quasiparticle pair
near the antinodal regime due to the residual superexchange interaction\cite{LeeRMP}. In this picture it would be hard to understand the necessity for the existence
of a singlet bound state with the roughly degenerate energy. The further
challenging and fundamental question is, given the presence of two degenerate
modes observed in ERS and INS, how can they directly influence the
superconducting coherence?

A proposal made by Uemura\cite{Uemura} recently is that the two
quasi-degenerate modes observed in INS and ERS may originate from soft modes in spin and charge channels in an incommensurate stripe state, which are called\cite{Uemura} \emph{twin spin/charge roton mode}, in analogy with the soft phonon-roton mode towards solidification in superfluid $^{4}\mathrm{He}$. Hence the mechanism for superconducting transition is due to the substantial
reduction of the superfluid density by thermal excitations of such twin
spin-charge soft mode at $T_{c}/2<T<T_{c}$, whereas the quasiparticle
excitations mainly dominate at lower temperature $<T_{c}/2$.

Nevertheless, according to the experimental results shown in Fig. \ref%
{fig:E-Tc}, it seems that the $T_{c}$ formula (\ref{eq:tcformula}) holds more
generally than simply in a neighborhood of stripe states\cite{stripe}. It calls for an
intrinsic \textquotedblleft spin-charge entanglement\textquotedblright \ in
the superconducting phase of the cuprates. Namely, magnetic excitations at $%
\mathbf{Q}_{\text{AF}}$ should have some kind of profound effect on the
superconducting condensation such that thermal excitations of the former can
be destructive to the latter, much more effective than the usual nodal
quasiparticles in the BCS theory. Furthermore, the mechanism should allow
for a degenerate singlet mode, which may be not associated with a soft mode
of any charge order as its characteristic momentum is around $\mathbf{Q}_{%
\text{0}}$, to play an equally important role. Lastly, the simple scaling
relation (\ref{eq:tcformula}) with a universal $\kappa $ should be independent
of the details of materials including the charge inhomogeneity. Or more
precisely, all the detailed properties of the system should influence $T_{c}$
mainly through the characteristic energy scale $E_{g}$.

\begin{figure}[t]
\includegraphics[width=4cm]{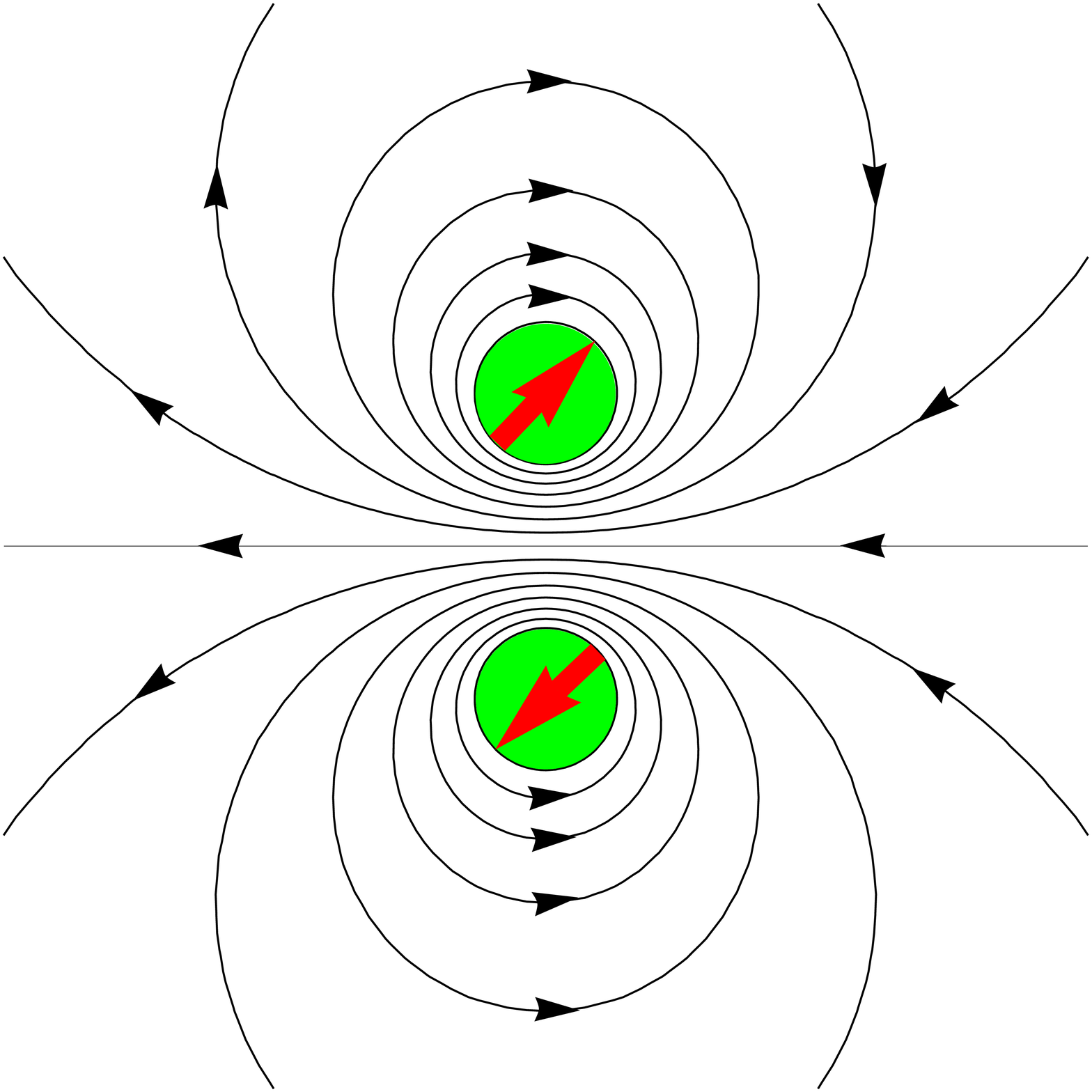}%
\includegraphics[width=4cm]{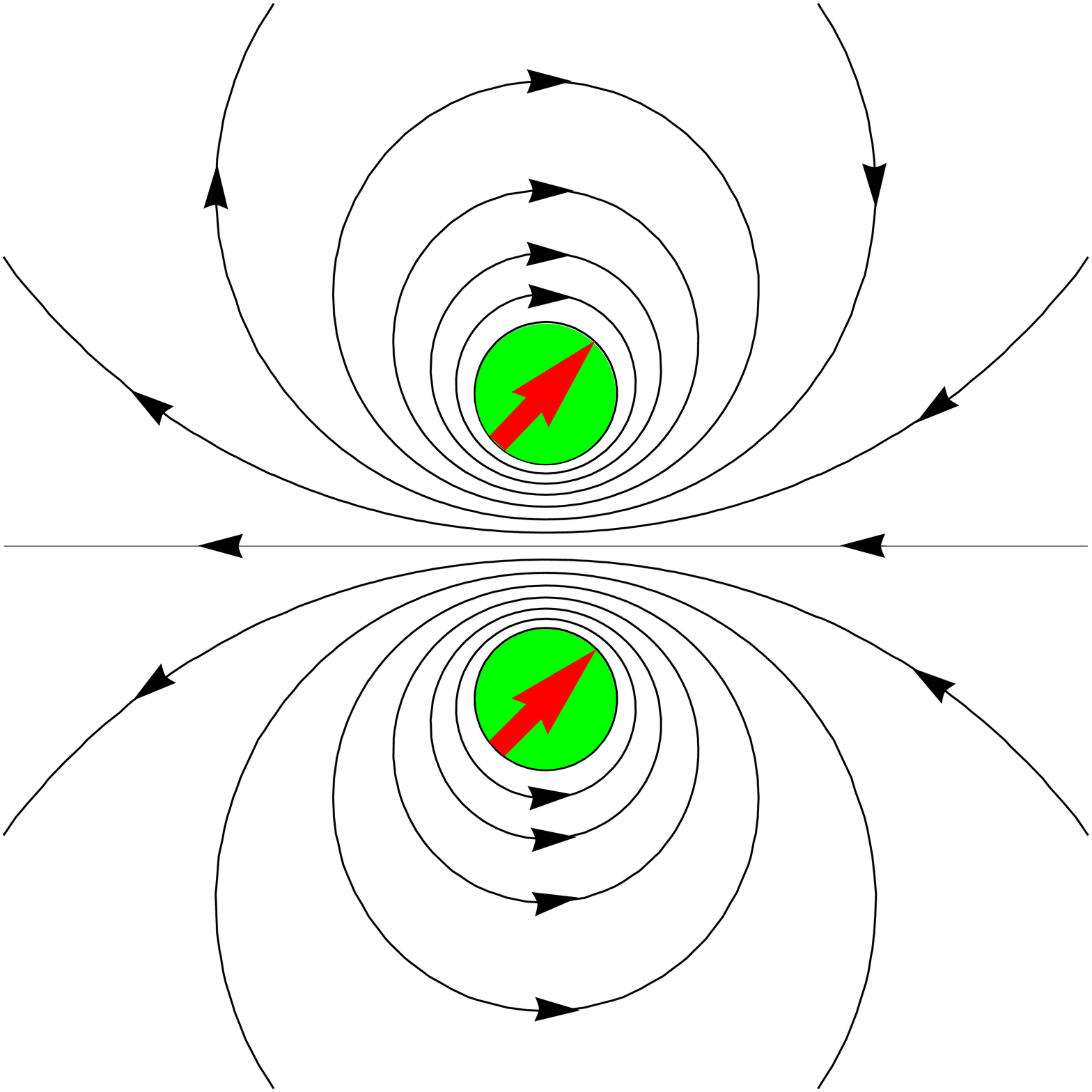}
\caption{[Color online] Schematic illustration of an $S=0$ (singlet) and an $S=1$ (triplet)
spin-rotons. Each of them is composed of a supercurrent vortex-antivortex
bound pair, with a pair of neutral free spins sitting at the two poles of
the two-dimensional  roton. Such a spin-roton composite is an elementary excitation in the
superconducting state of a doped Mott insulator described by the phase
string theory\cite{LPP_M}\cite{LPP_Q}. }
\label{fig:rotons}
\end{figure}

In this paper, we will demonstrate that a  self-consistent mathematical
description of superconductivity in doped Mott insulators can give rise to a systematic account for the above-mentioned novel properties including the $T_{c}$ formula (\ref
{eq:tcformula}). In the superconducting state, besides the emergent quasiparticles as the recombination of charge and spin, the most nontrivial elementary excitations are the vortex-antivortex bound
pairs locking with free spins at the poles, with total spin $S=0$ or $1$, as
illustrated in Fig. \ref{fig:rotons}. We shall call these excitations
\emph{spin-rotons} in the following, which are distinguished from those
proposed by Uemura\cite{Uemura} as they are not slaved with any charge and
spin orders, but a direct consequence of the phase string effect\cite{PhaseString,Rev_PhaseString} in the
$t$-$J$ model with a peculiar nonlocal spin-charge entanglement:
neutral spins locking with charge supercurrents\cite{LPP_M,LPP_Q}.

These spin-rotons will naturally include two degenerate excitations. The
degeneracy of these modes with spin quantum number $S=0$ and $1$ is due to
the fact that the pair of neutral spins are excited \textquotedblleft
spinons\textquotedblright \ from an underlying resonating-valence-bond
(RVB) spin background. The degenerate spin-roton modes thus indicate
spin-charge separation, but with a twist. That is, a stable spin-roton
object in the superconducting phase also implies a \emph{spinon-confinement} as two
spinons cannot be separated freely \emph{in space} due to the logarithmic potential
between the vortex and antivortex. Such rotonlike supercurrents will play a
central role in deciding the superconducting phase coherence transition as in (\ref{eq:tcformula}).
We will show that the singlet and triplet spin-rotons can be indeed
directly probed by ERS in \textrm{A}$_{1g}$ channel at $\mathbf{Q}_{\text{0}%
} $ and INS near $\mathbf{Q}_{\text{AF}}$. They have the minimal characteristic energy $E_{g}\sim \delta J$ in the low-doping regime with the magnitude in good agreement with
the experimental data where $\delta $ denotes the doping concentration and $%
J $ is the superexchange coupling constant determined in the undoped case.

The remainder of the paper is arranged as follows. In Sec. \ref{subsec:action}, we
introduce the description of a doped-Mott-insulator superconductor, obtained
previously\cite{LPP_M,LPP_Q} based on the phase string theory\cite{PhaseString,Rev_PhaseString} of the $t$-$J$ model, by using a
phenomenological construction. We argue that in order to incorporate the
influence of spin degrees of freedom (which is important in a lightly doped
Mott insulator where spins constitute the majority of low-lying degrees of
freedom) under the requirement of no time-reversal and spin rotational
symmetry breakings, one is naturally led to a modified action for superconductivity. In Sec. \ref{subsec:con}, the spin-roton excitations
as a direct consequence of this formulation are discussed. Then, how the spin-roton excitations as the resonancelike modes can be probed
by INS and ERS are discussed. In Sec. \ref{subsec:TC}, the $T_{c}$ formula (\ref%
{eq:tcformula}) determined by the spin-roton excitations is obtained. Finally,
in Sec. \ref{sec:diss}, a general discussion will be given.

\section{Spin-roton excitations\label{sec:roton}}

\subsection{Phenomenological description of a doped-Mott-insulator
superconductor\label{subsec:action}}

From a doped Mott insulator point of view\cite{Anderson}, the superconductivity in the
cuprates occurs in a small doping regime where the charge carrier number is
greatly reduced as compared to the total electron number. Namely, the strong
on-site Coulomb interaction will make the charge degrees of freedom
partially frozen, while the full spin degrees of freedom of the electrons
remain at low energy. Thus the London action (\ref{eq:london}) should be
modified in order to properly reflect the Mott physics.

For example, in the U(1) slave-boson gauge theory description\cite{Nagaosa}, the charge
carriers are described by spinless bosons known as holons. The
superconducting state corresponds to the Bose condensation of the holons,
with (\ref{eq:london}) replaced by
\begin{equation}
L_{h}=\frac{\rho _{s}}{2}\int d^{2}\mathbf{r}(\nabla \phi +\mathbf{A}^{s}+e%
\mathbf{A}^{e})^{2}  \label{eq:Lh}
\end{equation}%
where $\rho _{s}$ is proportional to the density of condensed holons and $%
q=+e$, in contrast to the conventional London action where the condensate of
Cooper pairs of the electrons is involved with $q=-2e$. As a component of
the electron fractionalization, holons are no longer gauge neutral and are
generally coupled to an internal emergent gauge field $\mathbf{A}^{s}$. In
the U(1) slave-boson gauge theory\cite{Nagaosa}, $\mathbf{A}^{s}$ will be also minimally
coupled to the other component of the electron fractionalization, i.e.,
neutral spins called spinons. However, since the latter are in RVB pairing,
the internal gauge field $\mathbf{A}^{s}$ is expected to be suppressed due
to the \textquotedblleft Meissner effect\textquotedblright \ of the RVB
state, whose mean-field transition temperature is presumably much higher at
low doping. Consequently in such a mean-field \textquotedblleft
pseudogap\textquotedblright \ regime $\mathbf{A}^{s}$ gains a mass and
cannot play a role as a new source to effectively reduce $T_{c}$\cite{Nagaosa}.

However, the U(1) slave-boson gauge theory is not the only possible
theoretical description for the doped Mott insulator. In the following, we
shall elucidate in a \emph{phenomenological} way an alternative self-consistent
construction. It will reveal the existence of a new mathematical structure\cite{LPP_M,LPP_Q},
in which the charge condensate can become strongly correlated with spin
excitations.

The key distinction will be that, instead of minimally coupling to both the
holon and spinon currents in the U(1) slave-boson gauge theory, here $
\mathbf{A}^{s}$ will only minimally couple to the holon matter field as
given in (\ref{eq:Lh}), not to spinon currents. Instead£¬ its strength will be
generated from the spinon matter field according to the following
gauge-invariant relation
\begin{equation}
\oint_{c}d\mathbf{r}\cdot \mathbf{A}^{s}(\mathbf{r})=\phi _{0}\int_{\Sigma
_{c}}d^{2}\mathbf{r}\left[ n_{\uparrow }^{b}(\mathbf{r})-n_{\downarrow }^{b}(\mathbf{r})\right] \label{eq:fluxAs}
\end{equation}%
Here the flux of $\mathbf{A}^{s}$ within an arbitrary loop $c$ on the
left-hand-side (l.h.s.) is contributed by $\pm \phi _{0}$ flux-tubes bound
to individual spinons on the right-hand-side (r.h.s.), with $n_{\uparrow
\downarrow }^{b}(\mathbf{r})$ denoting the local density of spinons where
the integration runs over the area $\Sigma _{c}$ enclosed by $c$.

Due to the sign change between the $\uparrow $ and $\downarrow $ spins on
the r.h.s. of (\ref{eq:fluxAs}), $\mathbf{A}^{s}(\mathbf{r})$ will explicitly
preserve the time-reversal (TR) symmetry, as $\uparrow $ $\leftrightarrow $ $%
\downarrow $ under the TR transformation. This is in contrast to a
conventional electromagnetic vector potential $\mathbf{A}^{e}$, which breaks
the TR symmetry. However, since the path $c$ is oriented, the spin
rotational symmetry may be broken for a general $\phi _{0}$. But under a
specific choice%
\begin{eqnarray}
\phi _{0}=\pi \label{eq:phi0}
\end{eqnarray}%
one finds that the spin rotational symmetry can be still maintained: without
loss of generality, one can consider a loop $c$ which encloses a single spin
such that $\oint_{c}d\mathbf{r}\cdot \mathbf{A}^{s}=\pm \phi _{0}=\pm \pi $
which is still spin-dependent. However, such a spin-dependence sign change can be
effectively compensated in (\ref{eq:Lh}) by combining with a proper topological
vortex of the holon condensate given in (\ref{eq:vor}). Such a \textquotedblleft
large\textquotedblright \ gauge transformation will not cost any energy in (%
\ref{eq:Lh}) when $\mathbf{A}^{e}=0$. It is also \textquotedblleft
legal\textquotedblright \ to precisely bind such a holon vortex core of (\ref%
{eq:vor}) with the spinon because the no double occupancy constraint in the
doped Mott insulator dictates that a site without a holon must be alway
occupied by a neutral spin.

Hence, based on some general physical considerations,
the London action for a superconducting state can be modified in a
fundamental way in a doped Mott insulator, with an internal vector potential
$\mathbf{A}^{s}$ emerging as a topological gauge field without breaking the
time-reversal and spin rotational symmetries.

\begin{figure}[t]
\includegraphics[width=3.5cm]{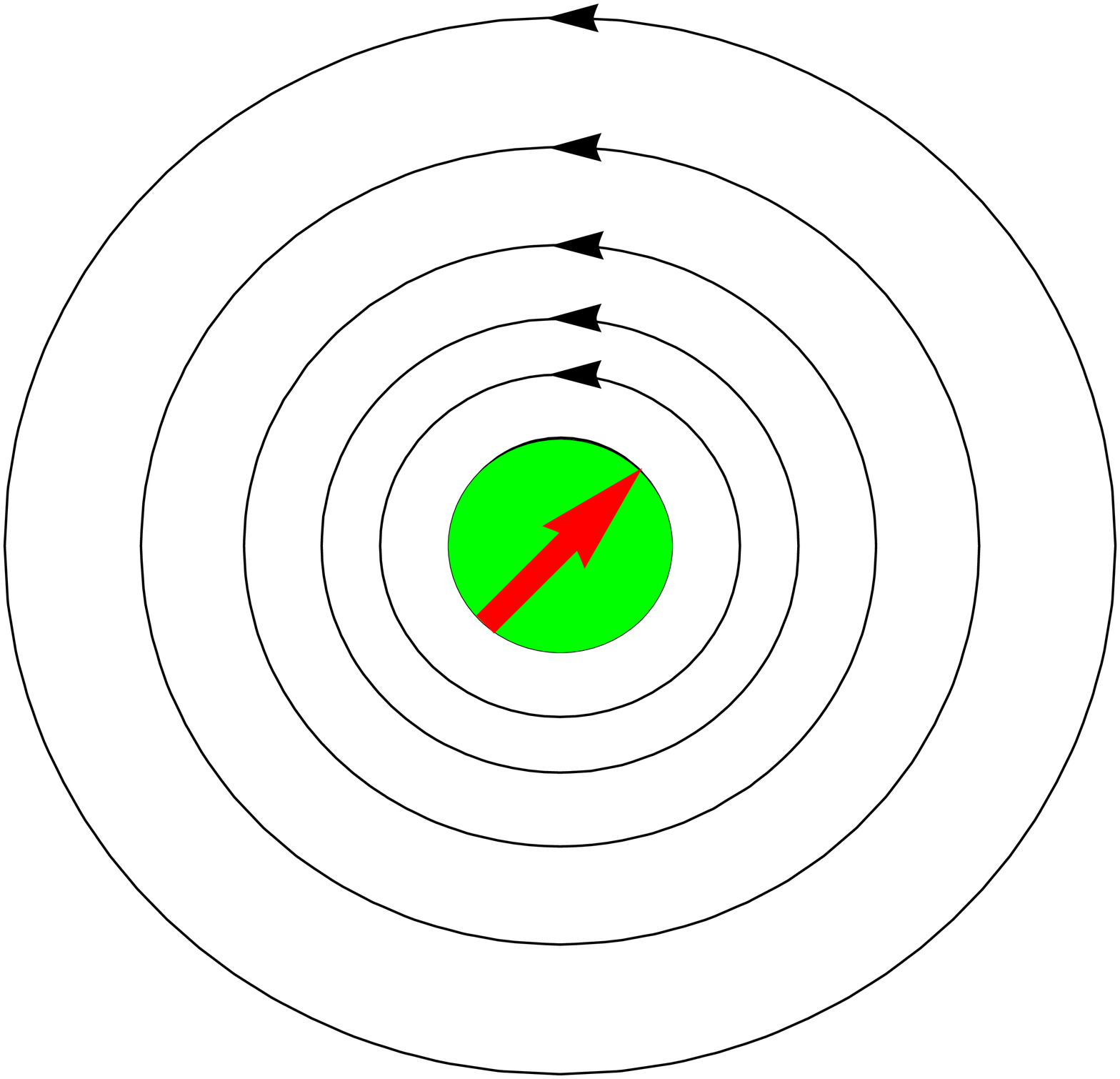} %
\includegraphics[width=3.5cm]{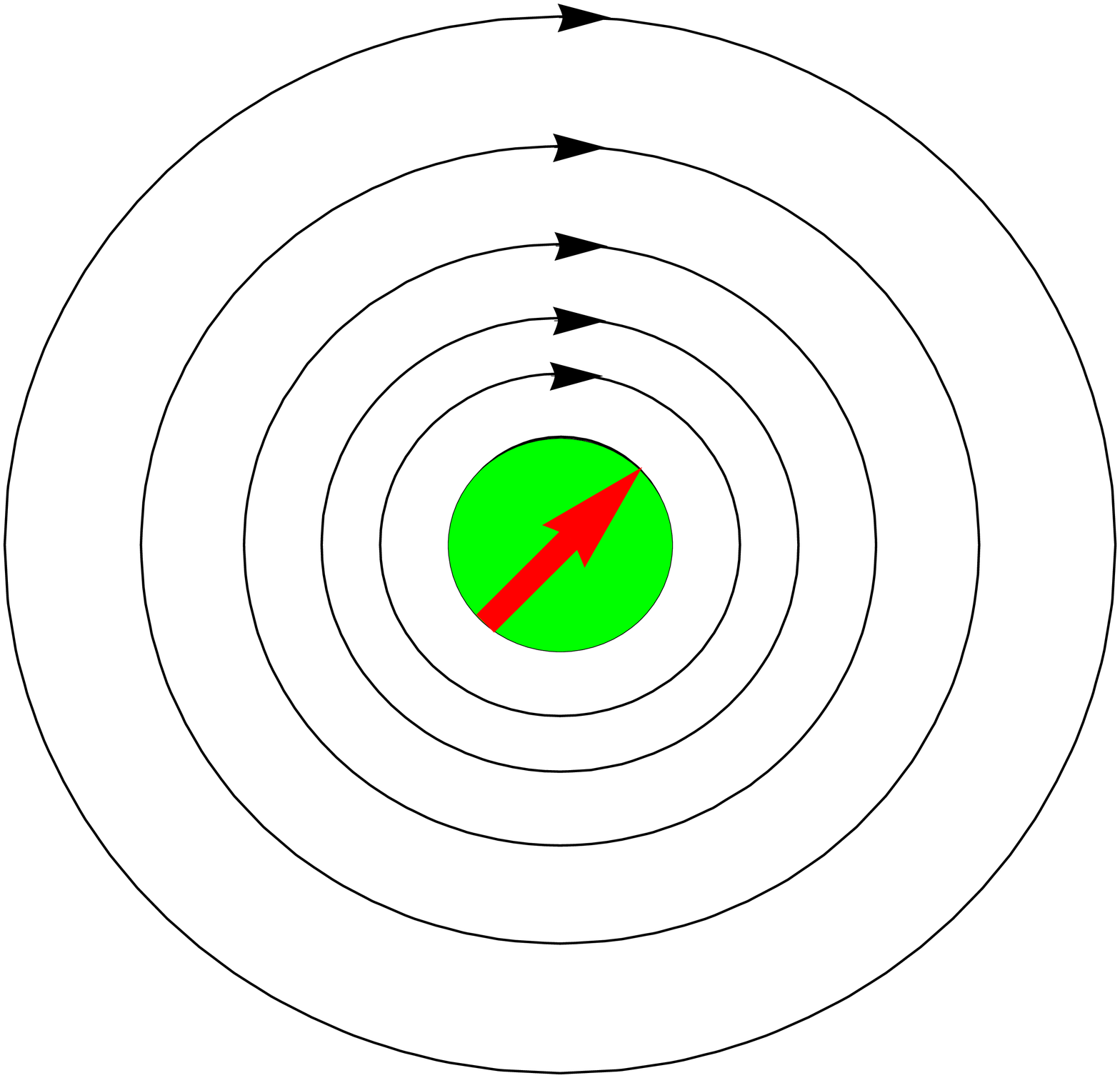}
\caption{[Color online] Schematic illustration of single spinon-vortices. An isolated
neutral spin (spinon) in the superconducting state will always induce a
vortexlike supercurrent response from the charge condensate according to the
generalized London action (\protect \ref{eq:Lh}). Notice that the vorticity sign
of the vortex is actually independent of the spin orientation as long as $%
\protect \phi _{0}=\protect \pi $ in (\protect \ref{eq:fluxAs}), which preserves the spin rotational symmetry. }
\label{fig:vortex}
\end{figure}

According to (\ref{eq:Lh}), the charge current will be determined by the London
equation%
\begin{eqnarray}
\mathbf{J}_{h}=\rho _{s}(\nabla \phi +\mathbf{A}^{s}+e\mathbf{A}^{e})\label{eq:le}
\end{eqnarray}%
For an isolated neutral spin, in terms of (\ref{eq:fluxAs}), there will be
vortexlike charge currents induced from the charge condensate with $\oint d\mathbf{r}\cdot \mathbf{J}_{h}=\pm \rho _{s}\pi $ in the absence of $\mathbf{A}^{e}$, where $\pm $ will be independent of the spin index based on the
above discussion. Namely each neutral spin can induce a current vortex with
two opposite vorticities as illustrated in Fig. \ref{fig:vortex}, which is
known as a spinon-vortex\cite{LPP_M,LPP_Q}.

According to a general argument given by Haldane and Wu\cite{Haldane}, since a spinon
behaves like a supercurrent vortex, its motion through a closed path $c$ must
then pick up a Berry's phase which is  determined by the number of superfluid
particles of the condensate in the area $\Sigma_c $ enclosed by $c,$ as if
it sees an effective \textquotedblleft magnetic-field\textquotedblright \
described by a vector potential $\mathbf{A}^{h}$:

\begin{eqnarray}
\Delta \Phi _{\mathrm{Berry}}(c) &=&\phi _{0}\int_{\Sigma _{c}}d^{2}\mathbf{r%
}~\rho _{h}(\mathbf{r})  \nonumber \\
&\equiv &\oint_{c}d\mathbf{r}\cdot \mathbf{A}^{h}(\mathbf{r})  \label{eq:fluxAh}
\end{eqnarray}%
Here $\rho _{h}(\mathbf{r})$ denotes the local superfluid density of
condensed holons, with $\phi _{0}=\pi $ instead of $2\pi $.

Thus, one may write down a minimal gauge-invariant Hamiltonian for spinons
simply as
\begin{equation}
H_{s}=-J_{s}\sum_{<ij>\sigma }b_{i\sigma }^{\dag }b_{j-\sigma }^{\dag
}e^{i\sigma A_{ij}^{h}}+\text{h.c.}  \label{eq:Hs}
\end{equation}%
where $b_{i\sigma }^{\dag }$ defines the bosonic creation operator for a
spinon at site $i$ with a spin index $\sigma $. Here $A_{ij}^{h}$ is the
lattice version of the gauge potential $\mathbf{A}^{h}(\mathbf{r})$
introduced in (\ref{eq:fluxAh}) and the sign $\sigma $ in front of the gauge
phase in (\ref{eq:Hs}) will ensure the TR invariance.

Although one can alternatively write down an effective model with the hopping
term $b_{i\sigma }^{\dag }b_{j\sigma }e^{i\sigma A_{ij}^{h}}$ replacing the
RVB pairing term $b_{i\sigma }^{\dag }b_{j-\sigma }^{\dag }e^{i\sigma
A_{ij}^{h}}$ in (\ref{eq:Hs}), without breaking the gauge and TR symmetries, (%
\ref{eq:Hs}) is physically more meaningful because in the ground state spinons
will be all paired up with $\left \langle b_{i\sigma }^{\dag }b_{j-\sigma
}^{\dag }e^{i\sigma A_{ij}^{h}}\right \rangle\equiv\Delta^s/2 \neq 0$ which automatically
satisfies the spinon-confinement requirement to ensure superconducting phase
coherence as to be discussed below. Furthermore, $\oint_{c}d\mathbf{r}\cdot
\mathbf{A}^{h}=0$ at half-filling, where $H_{s}$ (\ref{eq:Hs}) reduces to the
Schwinger-boson mean-field Hamiltonian which well captures the
antiferromagnetic (AF) correlations including the long-range AF order at $%
T=0$\cite{Arovas}.

Therefore, the London action (\ref{eq:london}) for superconductivity has been
phenomenologically modified  for the doped Mott insulator in (\ref{eq:Lh}). Here
the charge condensate will be generally coupled  to neutral spin excitations,
ubiquitously presented in a doped Mott insulator governed by (\ref{eq:Hs}),
via an emergent topological gauge field (\ref{eq:fluxAs}). Such a
self-consistent description based on (\ref{eq:Lh}), (\ref{eq:fluxAs}), (\ref{eq:fluxAh}) and (\ref{eq:Hs}) can be
justified\cite{LPP_M}\cite{LPP_Q} in the phase string theory of the $t$-$J$ model, with the
superfluid stiffness $\rho _{s}\equiv \rho _{h}/m_{h}$ ($m_{h}$ is the
effective mass for holons) and effective coupling constant $J_{s}$ in (\ref%
{eq:Hs}) determined microscopically. One is referred to Ref. \onlinecite{LPP_Q} and the
references therein for details. Although it is not a unique construction for
a doped Mott insulator (one can alternatively have other possible
mathematical constructions like the U(1) slave-boson gauge theory
description\cite{Nagaosa}, for example, as mentioned before), some very unique
consequences will follow from such a self-consistent approach, which can be
directly compared with experiments.

\subsection{Spin-roton excitations\label{subsec:con}}

A direct physical consequence is that a single spinon excitation in the
superconducting state will not be permitted because the self energy of a vortex shown in Fig. \ref{fig:vortex} is logarithmically divergent. Then all the spinons in the
superconducting state must appear in pairs, with the associated supercurrent
vortices forming vortex-antivortex bound pairs, as illustrated in Fig. \ref{fig:rotons}. These bound objects are referred to as spin-rotons, which
carry total spin $0$ (singlet) and $1$ (triplet), charge $0$, together with
a supercurrent structure analogous to a two-dimensional roton excitation in a Bose
condensate. In this sense, the spinons must be \textquotedblleft
confined\textquotedblright \ and only integer spin excitations are allowed
in the superconducting state.

\subsubsection{Resonancelike characteristic energy $E_{g}$\label{subsubsec:Eg}}

\begin{figure}[t]
\includegraphics[width=\columnwidth]{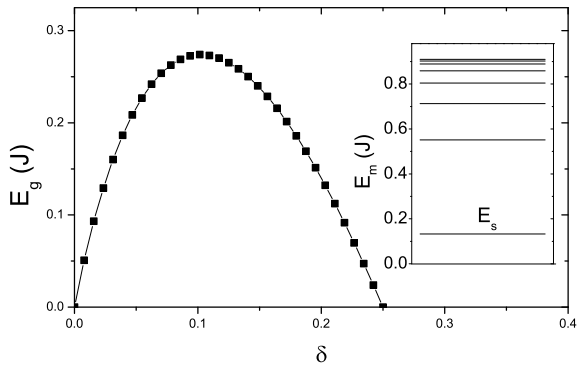}
\caption{The doping dependence of the characteristic energy scale $E_{g}$ of
the spin rotons is shown. Here $E_{g}=2E_{s}$ with $E_{s}$ as the lowest
excited energy of the spinon spectrum shown in the inset, obtained based on (%
\protect \ref{eq:Hs}) at a specific doping concentration $\protect \delta =0.125$%
. }
\label{fig:Eg}
\end{figure}

The spinon Hamiltonian (\ref{eq:Hs}) can be easily diagonalized \cite{Chen}
under the condition that the holons are uniformly condensed with $\rho
_{h}=\delta a^{-2}$ ($a$ is the lattice constant) as outlined in Appendix \ref{app:diag}. The solution of (\ref{eq:Hs}) has an uneven Landau-level-like spectrum for spinon excitations as shown in the inset of
Fig. \ref{fig:Eg}, which are excited by breaking up RVB pairs in the ground state.

\begin{figure}[b]
\includegraphics[width=5cm]{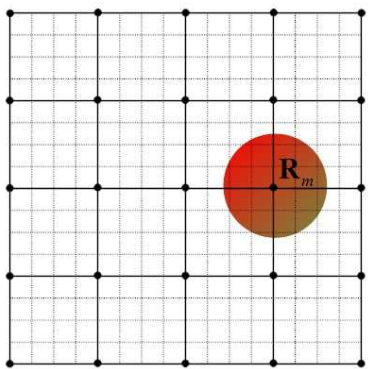}
\caption{[Color online] The degenerate spinon modes in the lowest energy level, shown in
the inset of Fig. \protect \ref{fig:Eg}, are labeld by $\mathbf{R}_{m}$ which
form the von Neumann lattice with a lattice constant $\protect \xi _{0}=%
\protect \sqrt{2\protect \pi }a_{c}$ with the cyclotron length $a_{c}=a/%
\protect \sqrt{\protect \pi \protect \delta }$ as the size of each spinon
wavepacket. Here the case $\protect \delta =1/8$ and $\protect \xi _{0}=4a$ is
shown. For each $\mathbf{R}_{m}$, there is an additional degeneracy $g=4$,
corresponding to orthogonal wavefunctions: $w_{m\uparrow }(\mathbf{r}_{i})$, $%
w_{m\downarrow }(\mathbf{r}_{i})$, $(-1)^{\mathbf{r}_{i}}w_{m\uparrow }(%
\mathbf{r}_{i})$ and $(-1)^{\mathbf{r}_{i}}w_{m\downarrow }(\mathbf{r}_{i})$%
\cite{Chen}. }
\label{fig:Neumann-lattice}
\end{figure}

At low temperature, we shall focus on the lowest excited level at $
E_{s}\equiv E_{g}/2$ for simplicity. In the main panel of Fig. \ref{fig:Eg},
$E_{g}=2E_{s}$ is shown as a function of doping under a proper consideration
 of the doping dependence of $J_{s}$\cite{Gu}. The corresponding spinon
wavepacket looks like
\begin{equation}
|w_{m\sigma }(\mathbf{r}_{i})|^{2}\simeq \frac{a^{2}}{2\pi a_{c}^{2}}\exp
\left \{ -\frac{\left \vert \mathbf{r}_{i}-\mathbf{R}_{m}\right \vert ^{2}}{%
2a_{c}^{2}}\right \} \label{eq:wp}
\end{equation}%
with a \textquotedblleft cyclotron length\textquotedblright \ $a_{c}\equiv a/%
\sqrt{\pi \delta }$. Namely, the lowest spinon excitations governed by (\ref%
{eq:Hs}) are non-propagating modes of an intrinsic size in order of $a_{c}$. Here
the degenerate levels are labeled by the coordinates $\mathbf{R%
}_{m}$\cite{Rashba}, the centers of the spinon wavepacket (\ref{eq:wp}), which form a von
Neumann lattice with a lattice constant $\xi _{0}=\sqrt{2\pi }a_{c}$, as
shown in Fig. \ref{fig:Neumann-lattice}.

After integration over the original lattice index $\mathbf{r}_{i}$ in the
modified London action (\ref{eq:Lh}) at $\mathbf{A}^{e}=0$, one can obtain (see
Appendix \ref{Mapping}) an effective interaction term for spinon-vortices on the von
Neumann lattice%
\begin{equation}
U_{\text{\textrm{int}}}=-\frac{\pi }{4}\rho _{s}\sum_{\mathbf{R}_m\mathbf{R}%
_{m^{\prime }}}\ln \frac{{|\mathbf{R}_{m}-\mathbf{R}_{m^{\prime }}|}}{\xi
_{0}}q_{m}q_{m^{\prime }}  \label{eq:Uint}
\end{equation}%
where $q_{m}$ ($=\pm 1$ or $0$) denotes the vorticity for each spinon-vortex on
the site $\mathbf{R}_{m}$, and to avoid the logarithmical divergence, the
neutral constraint $\sum_{m}q_{m}=0$ will be imposed. So the total energy of
the spinon-vortices is given by
\begin{equation}
H_{v}=\frac{E_{g}}{2}\sum_{m}|q_{m}|+U_{\text{\textrm{int}}}.  \label{eq:Hv}
\end{equation}%
It is noted that there is a four-fold degeneracy$,$ $g=4$, at each site $%
\mathbf{R}_{m}$ as mentioned in the caption of
Fig. \ref{fig:Neumann-lattice}.

Note that a conventional vortex-antivortex pair in a KT system will normally
shrink at low temperature and be annihilated in the ground state. But a spin-roton in the
present case cannot literally disappear in the ground state because the two
spins sitting at the poles of a roton in Fig. \ref{fig:rotons} cannot
annihilate each other. Nevertheless, the roton supercurrents surrounding the
neutral spins will have minimal effect on the
ground state. In fact, as the solution of (\ref{eq:Hs}), spins will form
short-range RVB pairs in the ground state, of a characteristic length scale $\sim a_c$
which is comparable to the finite core size of each pole of a spin-roton in
Fig. \ref{fig:rotons} (the spin trapped at the core cannot sit still due
to the uncertainty principle and the intrinsic core size is set by the cyclotron length $%
a_c$). Thus, the surrounding rotonlike supercurrents around an RVB pair
will be effectively canceled out in the ground state. In other words, the
London action (\ref{eq:Lh}) will be decoupled from the neutral RVB spin
background as $A^{s}\approx 0$ and the excited spinon-vortices are
effectively described by (\ref{eq:Hv}).

Hence the spin-roton structure shown in Fig. \ref{fig:rotons} will emerge as
the bound pair of the excited spinons, which are of spin triplet ($%
S=1) $ and singlet ($S=0)$, respectively, and \emph{degenerate} in energy.
The spin-rotons here will have a minimal energy scale $E_{g}$ when two
excited spinons are located at the same von Neumann lattice site such that
the vortex-antivortex supercurrent structure is effectively annihilated with
$U_{\text{\textrm{int}}}=0$.

The degenerate singlet and triplet spin-rotons imply the spin-charge
separation: i.e., the existence of single spinons carrying $S=1/2$ and zero
charge as individual excitations, which do not interact with each other
\emph{magnetically}. However, we have also seen that these spinons must be confined
\emph{spatially} in pairs, appearing at the poles of roton supercurrent structure
and subjected to logarithmic attraction $U_{\text{\textrm{int}}}$. Therefore,
in such a non-BCS superconducting state the spin-charge separation has a
twist, which is characterized by \emph{new} elementary excitations of
degenerate spin-rotons instead of individual spinons. In other words, the
spinon confinement does not mean a spin-charge tight recombination like in a
conventional Fermi liquid or BCS superconductor of the electrons. Rather, at
a short distance scale $\sim \xi _{0}$, the confining force $U_{\text{%
\textrm{int}}}$ becomes negligible and the spinons are still
\textquotedblleft asymptotically free\textquotedblright .

\subsubsection{INS and ERS probes\label{subsubsec:exp}}

Experimentally, the neutron and Raman scattering measurements can provide
direct means to probe such novel excitations, in spin triplet and singlet
channels, respectively.

Define the spin-spin correlation function
\begin{eqnarray}
\chi _{zz}(\tau ,\mathbf{r}_{i}-\mathbf{r}_{j})=-\langle T_{\tau
}S_{i}^{z}(\tau )S_{j}^{z}(0)\rangle
\end{eqnarray}
where $\tau $ denotes the imaginary time, $S_{i}^{z}=\frac{1}{2}\sum_{\sigma
}\sigma b_{i\sigma }^{\dag }b_{i\sigma }.$ Similarly a density-density
correlation function which can be detected by the electron Raman scattering%
\cite{Shastry} is defined as follows
\begin{eqnarray}
\chi _{\text{ERS}}=-\langle T_{\tau }\tau _{\mathrm{A}_{1g}}(\tau )\tau _{%
\mathrm{A}_{1g}}(0)\rangle
\end{eqnarray}
where the $\text{A}_{1g}$ density operator\cite{Shastry} $\tau _{\mathrm{A}_{1g}}\equiv -%
\frac{1}{2}\sum_{\left \langle ij\right \rangle \sigma }c_{i\sigma }^{\dag
}c_{j\sigma }+\mathrm{h.c}$. Here $c_{i\sigma }^{\dag }$ is the electron
operator whose relation with the holon and spinon operators is given in
Appendix \ref{app:diag}.

Based on the Bogolivbov transformation (\ref%
{bogo}) and the phase string representation for the electron operatore $c_{i\sigma}$, one can express $S_{i}^{z}$ and $c_{i\sigma }$ in terms of $\gamma
_{m\sigma }^{\dagger }$ and $\gamma _{m\sigma }$ as shown in Appendix A. We shall
mainly concentrate on energies near the minimal $E_{g}$, where the total
Hamiltonian reduces to $H_{v}$ (\ref{eq:Hv}) in which the interaction term $U_{%
\text{int}}$ can be also neglected because $S_{i}^{z}$ and $\tau _{\mathrm{A}%
_{1g}}$ only create a pair of spinons locally within a von Neumann lattice site
(Fig. \ref{fig:Neumann-lattice}):
\begin{equation}
S_{i}^{z}\sim -\frac{1}{2}\sum_{mn\sigma }u_{m}v_{n}w_{m\sigma }^{\ast }(\mathbf{r}%
_{i})w_{n\sigma }(\mathbf{r}_{i})\sigma \gamma _{m\sigma }^{\dag }\gamma
_{n-\sigma }^{\dag }+\text{h.c.}  \label{eq:Sz}
\end{equation}%
and
\begin{equation}
\tau _{A_{1g}}\sim -\delta \sum_{m\sigma }u_{m}|v_{m}|\gamma _{m\sigma
}^{\dag }\gamma _{m-\sigma }^{\dag }+\text{h.c.}  \label{eq:tau}
\end{equation}%
where $u_{m}v_{n}$ is the coherent factor due to the RVB paring, with $m$
and $n$ denoting the degenerate lowest energy states shown in the inset of Fig. \ref%
{fig:Eg} with the degenerate $E_{m}=E_s$.

\begin{figure}[t]
\includegraphics[width=\columnwidth]{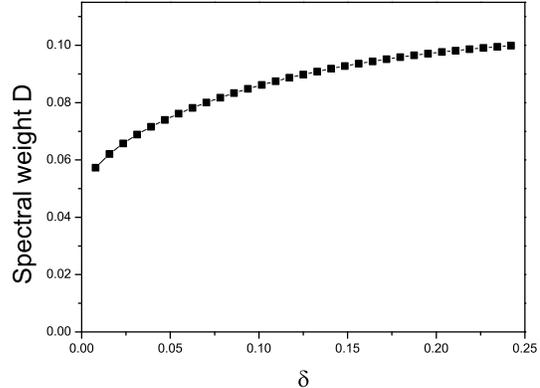}
\caption{The doping dependence of the spectral weight $D$ for the
spin-rotons appearing in (\protect \ref{eq:triplet}) and (\protect \ref{eq:singlet}%
). }
\label{fig:d}
\end{figure}

It is straightforward to obtain
\begin{eqnarray}
\chi _{zz}(\tau ,\mathbf{r})=-D(-1)^{\mathbf{r}}e^{-\frac{\mathbf{r}^{2}}{%
2a_{c}^{2}}}e^{-E_{g}\tau }
\end{eqnarray}
and
\begin{eqnarray}
\chi _{\text{ERS}}(\tau )\simeq -\delta De^{-E_{g}\tau }
\end{eqnarray}
with $D=\frac{\delta ^{2}}{2u_{m}^{2}v_{n}^{2}}$ is the spectral weight whose doping dependence is shown in Fig. \ref{fig:d}. In $\chi_{zz}$ we have used
the relation \cite{Rashba} $|\sum_{m}w_{m\sigma }^{\ast }(\mathbf{r})w_{m\sigma }(\mathbf{r%
}^{\prime })|=\frac{1}{2\pi a_{c}^{2}}e^{-\frac{(\mathbf{r}-\mathbf{%
r^{\prime }})^{2}}{4a_{c}^{2}}}$.

Correspondingly the dynamic spin susceptibility is obtained by
\begin{equation}
\chi _{\text{zz}}^{\prime \prime }(\mathbf{q},\omega )=\frac{2a_c^2}{\pi}D
e^{-2a_{c}^{2}(\mathbf{q}-\mathbf{Q}_{AF})^{2}}\delta (\omega -E_{g})
\label{eq:triplet}
\end{equation}%
and the $A_{1g}$ Raman scattering cross-section
\begin{equation}
I_{\text{ERS}}(\omega )\propto \chi _{\text{ERS}}^{\prime \prime }(\omega
)\simeq \delta D\delta (\omega -E_{g})  \label{eq:singlet}
\end{equation}
So the triplet spin-roton will appear as a resonancelike mode in $%
\chi _{\text{zz}}^{\prime \prime }(\mathbf{q},\omega )$ at $\omega =E_{g}$,
with momentum $\mathbf{q}$ peaked at the AF wavevector $\mathbf{Q}_{\mathrm{%
AF}}$ and a width inversely proportional to the
RVB pairing size $a_c$ which thus determines the spin-spin correlation
length $\propto a/\sqrt{\delta }$. Similarly, in $I_{\text{ERS}}(\omega )$ a
\textquotedblleft resonance mode\textquotedblright \ at $E_{g}$ will also be
exhibited, which corresponds to the singlet spin-roton excitation. It should be
emphasized that in the neutron and Raman scattering measurements only local spinons at the same von Neumann lattice are involved and the correction from the logarithmic potential $U_{%
\mathrm{int}}$ in (\ref{eq:Hv}) is always negligible. Of course, high-energy
spin-roton excitations can be also detected by these experiments at $\omega
>E_{g}$, which will involve spinons at higher energy levels shown in the
inset of Fig. \ref{fig:Eg}, whose effect\cite{Chen} will not be considered in the
present work for simplicity.

At half-filling, the minimal roton energy will be softened to zero: i.e., $%
E_{g}=0$ with $a_{c}\rightarrow \infty $. As shown in Fig. \ref{fig:d} the spectral weight  $D$ in (\ref{eq:triplet}) remains finite at $\delta \rightarrow 0$ and
characterizes the weight of the N\'{e}el oder as the triplet rotons at $%
E_{g}=0$ are condensed into the AF ordering. By contrast, $I_{\text{ERS}%
}(\omega )=0$ in this limit as there is no  more charge density fluctuation
to couple with the incident light in the Raman scattering measurement.
Furthermore, high-energy triplet spin-roton excitation is expected to be reduced
to the gapless spin wave\cite{Chen} at $\delta \rightarrow 0$ with the
spinon spectrum shown in the inset of Fig. \ref{fig:Eg} becomes a continuous
energy spectrum described by the Schwinger boson mean-field theory\cite{Arovas}.

\subsection{$T_{c}$ formula \label{subsec:TC}}

We now discuss how thermally excited spin-rotons can effectively destroy the
phase coherence of the superconducting condensation and determine the
transition temperature $T_{c}$.

The long-wavelength superfluid stiffness $\rho _{s}$ will be renormalized by
spin-roton excitations via the internal gauge field $\mathbf{A}^{s}$ in the
London action (\ref{eq:Lh}). Such spin-rotons shown in Fig. \ref{fig:rotons}
resemble the conventional vortex-antivortex pairs in the $XY$ model\cite{Kosterlitz}, except that
the unit vorticity of each spinon-vortex is $\pi $ instead of $2\pi $ of a
conventional vortex. A further difference is that the low-energy
spinon-vortices will distribute on a von Neumann lattice with the degeneracy $g=4$  as illustrated in
Fig. \ref{fig:Neumann-lattice}, instead of $g=1$ on the orginal lattice in the $XY$
model. Corresponding to the minimal energy $E_{g}$ of a spin-roton,  the fugacity is $y=e^{-E_{g}/2k_BT}$ as each spinon effectively contributes
to a core energy $E_{g}/2$. Compared to the $XY$ model, such a vortex core
energy is much cheaper as $E_{g}\sim \delta J$ at low doping. Thus,
the superconducting phase transition controlled by spin-rotons, which are
governed by $H_{v}$ in (\ref{eq:Hv}), is expected to be similar to a
conventional KT transition, but the $T_{c}$ formula should be quantitatively
different due to the peculiar internal structure of a spin-roton excitation
outlined above.

In the following, we shall follow a standard textbook mathematic procedure\cite{Chaikin} in dealing
with a conventional KT transition. Define the
reduced stiffness $K\equiv \frac{\rho _{s}}{k_BT}.$ and then the renormalized
reduced stiffness $K_{R}$, obtained by averaging over the spin-roton
excitations governed by (\ref{eq:Hv}), is found by
\begin{eqnarray}
K_{R}=K+\frac{\pi ^{2}K^{2}}{4Na^{2}}g^2\sum_{\mathbf{R}_{m}\mathbf{R}%
_{m^{\prime }}}(\mathbf{R}_{m}-\mathbf{R}_{m^{\prime }})^{2}\langle
q_{m}q_{m^{\prime }}\rangle
\end{eqnarray}%
where $N$ is the original total lattice number. The correlation $\langle q_{m}q_{m^{\prime }}\rangle $ can be easily
evaluated in terms of (\ref{eq:Hv}) to lowest order in fugacity $y$\cite{Chaikin}:
\begin{eqnarray}
\langle q_{m}q_{m^{\prime }}\rangle =-2y^{2}\left[ \left \vert \mathbf{R%
}_{m}-\mathbf{R}_{m^{\prime }}\right \vert /\xi _{0}\right] ^{-\frac{\pi }{2}%
K}.
\end{eqnarray}
such that
\begin{equation}
K_{R}=K-g^{2}\pi ^{3}y^{2}K^{2}\int_{\xi _{0}}^{\infty }\frac{dR}{\xi _{0}}%
\left( \frac{R}{\xi _{0}}\right) ^{3-\frac{\pi }{2}K}  \label{eq:KR}
\end{equation}%
where the lattice constant $\xi _{0}$ of the von Neumann lattice provides
the short distance cutoff. Again following the steps in Ref.  \onlinecite{Chaikin},
one arrives at differential renormalization group (RG) equations%
\begin{eqnarray}
&&\frac{dK^{-1}}{dl} =g^{2}\pi ^{3}y^{2}+O(y^{4})  \\
&&\frac{dy}{dl} =\left( 2-\frac{\pi }{4}K\right) y+O(y^3)
\end{eqnarray}%
with $K_{R}=K_{R}[K(l),y(l)]$ remaining as a constant, which results in the
fixed point at $K^{\ast }=8/\pi $ and $y^{\ast }=0$.
\begin{figure}[t]
\includegraphics[width=\columnwidth]{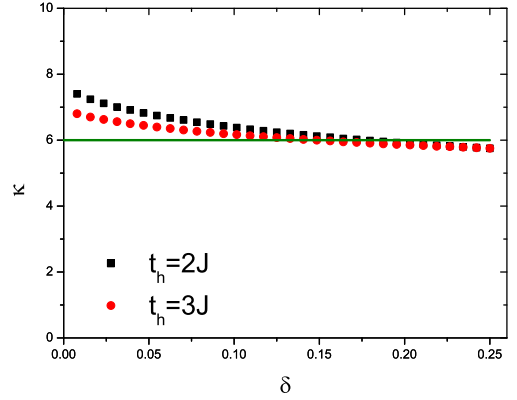}
\caption{[Color online] The coefficient $\protect \kappa $ defined in (\protect \ref{eq:Tc}) is
calculated at some typical values of the parameter: $t_{h}/J=2$ and $3$, and is weakly doping dependent around $\kappa =6$.}
\label{fig:K}
\end{figure}

Thus, by substituting $K_{R}=\lim_{l\rightarrow \infty }K(l)=K^{\ast }$ into
(\ref{eq:KR}), one gets
\begin{eqnarray}
\frac{8}{\pi }=\frac{\rho _{s}}{k_BT_{c}}+g^{2}\pi ^{3}y_c^{2}\frac{\left( \frac{%
\rho _{s}}{k_BT_{c}}\right) ^{2}}{4-\frac{\pi }{2}\frac{\rho _{s}}{k_BT_{c}}}
\end{eqnarray}
with $y_c^2=e^{-E_g/k_BT_c}$, which can be further rewritten as
\begin{eqnarray}
  y_c^2=\frac{1}{2\pi^2}\frac{n^2}{g^2}(1-\frac{8}{n^2\pi}\frac{k_BT_c}{\rho_s})^2
  \label{eq:pairn}
\end{eqnarray}
in which $n\pi$ with $n=1$ denotes the unit vorticity of the vortex.
(For the sake of comparison, we have introduced $n$ in
(\ref{eq:pairn}) such that the case of $n=2$ is also allowed which
corresponds to the conventional $2\pi$ vortex in the $XY$ model.)
Equation (\ref{eq:pairn}) indicates that the rigidity of the
superconducting state  can only sustain the amount of
vortex-antivortex pairs with
$y_c^2\leq\frac{1}{2\pi^2}\frac{n^2}{g^2}$. Using $n=1$ and $g=4$,
one finally finds
\begin{eqnarray}
\frac{E_{g}}{k_BT_{c}}=2\ln \frac{4\sqrt{2}\pi }{1-\frac{8k_BT_{c}}{\pi \rho _{s}}}%
\equiv \kappa \label{eq:Tc}
\end{eqnarray}%
which at $k_BT_{c}\ll $ $\pi \rho _{s}/8$ results in
\begin{eqnarray}
\kappa \simeq 2\ln 4\sqrt{2}\pi =5.76
\end{eqnarray}
Generally, $\kappa $ can be determined self-consistently according to (\ref%
{eq:Tc}) with using $m_{h}=1/(2t_{h}a^{2})$ and $E_{g}(\delta )$ presented in
Fig. \ref{fig:Eg}. The result is shown in Fig. \ref{fig:K} as a function of
doping concentration $\delta $ at $t_{h}=2J$ and $t_{h}=3J$, respectively.
Fig. \ref{fig:K} indicates that the value of $\kappa $ is roughly a
universal value at $6$ which is weakly dependent on the choice of $%
t_{h}$ as well as the doping concentration. So we obtain the $T_{c}$ formula (%
\ref{eq:tcformula}), which is in excellent agreement with the high-$T_c$ cuprates
as shown by the straightline in Fig. \ref{fig:E-Tc}. It is noted that $%
y_{c}^{2}=e^{-\kappa }\ll 1$ is consistent
with the small fugacity condition used in the above derivation of the RG
equations. Finally, we comment that in a previous more complicated approach\cite{Shaw},
$T_c$ was calculated without properly considering both the singlet and triplet spin-roton excitations,
which resulted in somewhat higher and non-universal value of $T_c$.

\section{Discussion\label{sec:diss}}

In this work, we have proposed a consistent understanding of some intriguing
experimental facts concerning high-$T_{c}$ superconductivity in the
cuprates. The key concept is the presence of a new type of elementary
excitations in the superconducting state, \textrm{i.e., }spin-rotons, in
addition to conventional nodal quasiparticle excitations. Such novel modes
are composed of supercurrent vortex-antivortex pairs (rotons) locking with
free spins at the two poles, which form degenerate spin singlet and triplet
spin states. We have found that they are indeed measurable by ERS in the $
A_{1g}$ channel and INS at the AF wavevector as resonancelike
modes, which are consistent with the experimental observations. In
particular, we have shown that it is this new kind of excitation that
determines superconducting phase coherence transition with $T_{c}\propto
E_{g}$ in (\ref{eq:tcformula}), in excellent agreement with the cuprate
superconductors. It should be noted that the $A_{1g}$ peak has also been probed
in the resonant electronic Raman scattering experiment\cite{RERS}. Similarly,
the resonant inelastic X-ray scattering (RIXS)\cite{RIXS} should be also able to detect
such a singlet spin-roton excitation if a higher resolution ($\leq40\text{emV}$) can be achieved.

So the \textquotedblleft resonance energy\textquotedblright $E_g$ as the characteristic energy scale
of these spin-roton excitations will play an important role in the superconducting phase, in contrast
to the BCS theory in which quasiparticle excitations dominate. To leading order approximation, $E_g$ vs.
doping in Fig. \ref{fig:Eg} will decide the phase diagram of superconductivity. Here $E_g$ (thus $T_c$)
vanishes at overdoping because the underlying RVB pairing
$\Delta^s\equiv\sum_\sigma\langle b_{i\sigma}^\dag b_{j-\sigma}^\dag e^{i\sigma A_{ij}^h}\rangle=0$\cite{Gu},
while $E_g$ vanishes at $\delta=0$ where the spin-rotons experience Bose condensation to form an AF N\'{e}el
order at $T=0$. The phase above $T_c$ will be full of free spinon-vortices  known as the spontaneous vortex
phase or the lower pseudogap phase\cite{LPP_M,LPP_Q}, which may explain the Nernst regime discovered\cite{Nernst}
in the cuprates.

However, if $E_g$ vanishes at a finite but small doping concentration, then the AF order may persist over in a
finite regime where $T_c=0$. As a matter of fact, if the non-uniform charge distribution is allowed, $E_g$ as the
solution of (\ref{eq:Hs}) can indeed be softened to zero at some small finite doping. A case considering some $Z_2$
topological excitation at low doping does lead to the result that $E_g$ vanishes as $\sqrt{\delta-x_c}$ at a critical
doping $x_c\simeq0.043$\cite{Kou}. Below $x_c$, either an AF spin glass state or charge stripe phases has been shown\cite{Kou,Kou_b}
to be competitive before the system becomes a commensurate AF ordered state near the half-filling.

Furthermore, there is no physical reason to protect the degeneracy of
singlet and triplet spin-roton excitations as $E_{g}\rightarrow 0$. In other
words, the residual interaction may decide which mode will be softened more
quickly to result in a competing charge or spin order at low doping, as
conjectured in Refs. \onlinecite{Uemura} and \onlinecite{LDH}. The bottomline here is that the spin-roton excitations are
expected to be essential in describing the quantum phase transition  of superconductivity to other low-doping phases at $T=0$.
Detail investigation along this line is beyond our current scope and will be discussed elsewhere.

\begin{acknowledgments}
We acknowledge helpful discussions with P. W. Anderson, D. H. Lee,
P. A. Lee, T. Li, N. P. Ong, T. Senthil, and X. G. Wen. This work is
supported by the NSFC grant nos. 10688401, 10834003 and the National
Program for Basic Research of MOST grant no. 2009CB929402.
\end{acknowledgments}
\appendix

\section{Diagonalizations of $H_s$ (\ref{eq:Hs})\label{app:diag}}
The spinon Hamiltonian $H_s$ (\ref{eq:Hs}) can be easily diagonalized\cite{Chen} under a uniform distribution of the holon condensate
$\rho_h=\delta a^{-2}$. To be self-contained, in the following we briefly outline the main results.

By using the Bogoliubov transformation
\begin{equation}
b_{\sigma }(\mathbf{r})=\sum_{m}(u_{m}\gamma _{m\sigma }-v_{m}\gamma
_{m-\sigma }^{\dag })w_{m\sigma }(\mathbf{r})  \label{bogo}
\end{equation}%
we obtain the spinon Hamiltonian $H_s$ as follows
\begin{eqnarray}
H_{s}=\sum_{m\sigma }E_{m}\gamma _{m\sigma }^{\dag }\gamma _{m\sigma }+\text{
const.}
\end{eqnarray}
where
\begin{eqnarray}
u_m=\frac{1}{\sqrt{2}}\sqrt{\frac{\lambda}{E_m}+1},~~ v_m=\text{sgn}(\xi_m)%
\frac{1}{\sqrt{2}}\sqrt{\frac{\lambda}{E_m}-1}
\end{eqnarray}
and
\begin{eqnarray}
E_{m} =\sqrt{\lambda ^{2}-\epsilon _{m}^{2}}\label{eq:Spectum}
\end{eqnarray}
Here the quantum number $m$ denotes an eigen-state $w_{m\sigma}(\mathbf{r}_i)$ with the eigen-value $\xi_m$
\begin{eqnarray}
\epsilon _{m}w_{m\sigma }(\mathbf{r}_{i}) =-J_{s}\sum_{j=\text{NN}%
(i)}e^{i\sigma A_{ij}^{h}}w_{m\sigma }(\mathbf{r}_{j})  \label{eq:w}
\end{eqnarray}
The spinon excitation spectrum (\ref{eq:Spectum}) exhibits an uneven
Landau-level-like form as shown in the inset of Fig. \ref{fig:Eg}.
To obtain this spectrum, we have used a self-consistent condition
for $J_s=J(1-4\delta)\Delta^s/2$ \cite{Gu} and the chemical
potential $\lambda$ in $E_m$ by enforcing
$\sum_{i\sigma}b_{i\sigma}^\dag b_{i\sigma}=(1-\delta)N$.

Focusing on the lowest energy level $E_s=E_g/2$, the corresponding wave package as the solution of (\ref{eq:w}) can be express as
\begin{eqnarray}
  O_{m}(\mathbf{r}_{i})=|w_{m\sigma }(\mathbf{r}_{i})|^{2}\simeq \frac{a^{2}}{%
2\pi a_{c}^{2}}\exp \left \{ -\frac{1}{2a_{c}^{2}}|\mathbf{r}_{i}-\mathbf{R}%
_{m}|^2\right \}\nonumber\\
\label{eq:O}
\end{eqnarray}
with the degenerate states labeled\cite{Rashba} by the site $\mathbf{R}_m$ in a von Neumann lattice shown in Fig. \ref{fig:Neumann-lattice}.
Note that for each $\mathbf{R}_m$, there are four degenrate states (g=4) corresponding $w_{m\uparrow}(\mathbf{r}_i)$, $w_{m\downarrow}(\mathbf{r}_i)$,
$(-1)^{\mathbf{r}_i}w_{m\uparrow}(\mathbf{r}_i)$ and $(-1)^{\mathbf{r}_i}w_{m\downarrow}(\mathbf{r}_i)$ due to the time reversal and bipartite lattice symmetry\cite{Chen}.

Finally one can express  the spin operator $S_i^z=\frac{1}{2}\sum_\sigma\sigma b_{i\sigma}^\dag b_{i\sigma}$ in terms of $\gamma_{m\sigma}^\dag$ and $\gamma_{m\sigma}$
\begin{eqnarray}
  S^z_i&=&\frac{1}{2}\sum_{mn\sigma}\sigma(u_m\gamma_{m\sigma}^\dag-v_m\gamma_{m-\sigma})
  (u_n\gamma_{n\sigma}-v_n\gamma_{n-\sigma}^\dag)\nonumber\\
  &&w^*_{m\sigma}(\mathbf{r}_i)w_{n\sigma}(\mathbf{r}_i)\nonumber\\
  &\simeq&-\frac{1}{2}\sum_{mn\sigma}\sigma u_mv_nw^*_{m\sigma}(\mathbf{r}_i)w_{n\sigma}(\mathbf{r}_i)\gamma^\dag_{m\sigma}\gamma^\dag_{n-\sigma}
  +\text{h.c.}\nonumber\\
\end{eqnarray}
Here we discard the $\gamma\gamma$ terms because they have vanishing contribution at the low temperature. And for the Raman tensor in the $A_{1g}$ channel\cite{Shastry}, $\tau_{A_{1g}}\equiv-\frac{1}{2}\sum_{<ij>\sigma}c_{i\sigma}^\dag c_{j\sigma}+\text{h.c.}$, one can use the phase string representation\cite{PhaseString,Rev_PhaseString} for the electron operator in the $t$-$J$ model and the holon condensation condition to obtain
\begin{eqnarray}
  \tau_{A_{1g}}
  &=&-\frac{1}{2}\sum_{<ij>\sigma}h_i h_j^\dag e^{-i(\phi_{ij}^0+A_{ij}^s)}b_{i\sigma}^\dag b_{j\sigma}e^{i\sigma A_{ij}^h}+\text{h.c.}\nonumber\\
  &\simeq&-\frac{1}{2}\delta\sum_{<ij>\sigma}b_{i\sigma}^\dag b_{j\sigma}e^{i\sigma A_{ij}^h}+\text{h.c.}\nonumber\\
  &\propto&-\delta\sum_{m\sigma}u_m|v_m|\gamma^\dag_{m\sigma}\gamma^\dag_{m-\sigma}+\text{h.c.}
\end{eqnarray}

\section{Derivation of $U_{int}$ in (\ref{eq:Uint})\label{Mapping}}
According to the discussion in Sec. \ref{subsec:action}, an excited spinon will always induce a $\pi$-vortex as shown in Fig. \ref{fig:vortex}. The vortex core will be determined by the spinon wavepacket $O_m(\mathbf{r}_i)$ in (\ref{eq:O}) with a core energy $E_s=E_g/2$.  Introduce $\mathbf{m}=\nabla \times \tilde{\mathbf{A}}$ with $\tilde{\mathbf{A}}\equiv \nabla\phi+\mathbf{A}^s$ to describe the winding number for the spinon vortices:
\begin{eqnarray}
\mathbf{m}(\mathbf{r})=\hat{z}\pi \sum_{m}\sum_{\mathbf{r}_{i}}O_{m}(\mathbf{%
r}_{i})\delta (\mathbf{r}-\mathbf{r}_{i})q_{m}
\end{eqnarray}
where $q_m$($=0,\pm1$) denotes the vorticity of a spinon-vortex ($q_m=0$ means no spinon excitation at state $m$). Then by integrating over $\mathbf{r}$ in (\ref{eq:Lh}) in the absence of the external electromagnetic field, one can determine an effective interaction between the spinon-vortices
\begin{eqnarray}
U_{\text{int}} &=&\frac{1}{2}\rho _{s}\int \frac{d^{2}\mathbf{q}}{(2\pi )^{2}} \tilde{%
\mathbf{A}}(\mathbf{q})\cdot \tilde{\mathbf{A}}(-\mathbf{q})  \nonumber \\
&=&\frac{1}{2}\rho _{s}\int \frac{d^{2}\mathbf{q}}{(2\pi )^{2}}\frac{\mathbf{m}(%
\mathbf{q})\cdot \mathbf{m}(-\mathbf{q})}{q^{2}}  \nonumber \\
&=&Q^{2}\frac{\pi }{4}\rho _{s}\ln L-\frac{\pi }{4}\rho _{s}\sum_{\mathbf{R}%
_m\mathbf{R}_{m^{\prime }}}q_{m}q_{m^{\prime }}I_{mm^{\prime }}
\end{eqnarray}%
The first term in $U_{\text{\textrm{int}}}$ leads to vortex neutrality $%
Q=\sum_{m}q_{m}=0$, and in the second term
\begin{eqnarray}
I_{mm^{\prime }} &=&\sum_{i,j}O_{m}(\mathbf{r}_{i})O_{m^{\prime }}(\mathbf{r}%
_{j})\ln |\mathbf{r}_{i}-\mathbf{r}_{j}|  \nonumber \\
&=&\sum_{\mathbf{r}_{i}',\mathbf{r}_{j}'}\left( \frac{a^{2}}{2\pi a_{c}^{2}}\right) \exp \left \{ -\frac{%
a^{2}}{2a_{c}^{2}}(\mathbf{r}_{i}'^{2}+\mathbf{r}_{j}'^{2})\right \}  \nonumber
\\
&&\times \ln |(\mathbf{r}_{i}'-\mathbf{r}_{j}')+(\mathbf{R}_{m}-\mathbf{R}%
_{m^{\prime }})|  \nonumber \\
&\simeq &\ln |(\mathbf{R}_{m}-\mathbf{R}_{m^{\prime }})|
\end{eqnarray}%
which leads to (\ref{eq:Uint}).

\end{document}